\documentclass{appolb}
\usepackage{epsfig}
\usepackage{graphicx}
\usepackage{amssymb}
\usepackage{dcolumn}
\usepackage{amsmath}
\usepackage{bm}
\usepackage{textcomp}

\begin{document}
\title{Static critical fluctuations on the freeze-out surface
\thanks{This work is supported by the NSFC and the MOST under grant
Nos. 11435001 and 2015CB856900.}%
}

\author{Lijia Jiang,
\address{Frankfurt Institute for Advanced Studies, Ruth-Moufang-Strasse 1, 60438 Frankfurt am Main, Germany}
\address{Department of Physics and State Key Laboratory of Nuclear Physics and Technology, Peking University, Beijing 100871, China}
\and
Pengfei Li,
\address{Department of Physics and State Key Laboratory of Nuclear Physics and
Technology, Peking University, Beijing 100871, China}
\and
Huichao Song
\address{Department of Physics and State Key Laboratory of Nuclear Physics and
Technology, Peking University, Beijing 100871, China}
\address{Collaborative Innovation Center of Quantum Matter, Beijing 100871, China}
\address{Center for High Energy Physics, Peking University, Beijing 100871, China}
}
\maketitle

\begin{abstract}
In this proceeding, we summarize the main results of our recent paper,
which introduces a freeze-out scheme to the dynamical models near the QCD critical point.
Within such framework of static critical fluctuations, the Beam Energy Scan (BES) data of $C_4$ and $\kappa \sigma^2$ for net protons within different $p_T$ ranges can be roughly described. Besides, the momentum acceptance dependence of  higher cumulants at lower collision energies can also be qualitatively described. However, $C_2$ and $C_3$ are always over-predicted due to the positive static critical fluctuations.
\end{abstract}
\section{Introduction}
The QCD critical point is the landmark point on the QCD phase diagram. The divergence of the correlation length in the vicinity of the critical point leads to the divergence of particle multiplicity fluctuations, which provides an accessible way in experiments to detect the existence of the critical point~\cite{Stephanov:2008qz,Stephanov:2011pb}. The recent BES data of $\kappa \sigma^2$ presented large deviations from the Poisson baselines at lower collision energies, and non-monotonic behavior at around $20$ GeV~\cite{Aggarwal:2010wy, Adamczyk:2013dal,Luo:2015ewa}, which indicates possible signals for the existence of the QCD critical point. Dynamical models have been built to describe the evolution of bulk matter and the chiral field in heavy ion collisions~\cite{Paech:2003fe}. However, a proper treatment of the freeze-out scheme is lacked. In an earlier paper \cite{Jiang}, we introduce a freeze-out scheme to such dynamical models, which includes the critical fluctuations near the critical point. The main results are summarized in this proceeding.

\section{The model and set ups}
Our calculations are based on the assumption that the distribution of protons emitted from the fireball near the critical point satisfies the static statistics, but with a variable effective mass, which is induced by the interaction between the $\sigma$ field and protons. The fluctuations of the $\sigma$ field transfer to the fluctuations of protons mass, leading to the fluctuations of proton distributions.
Such critical fluctuations can be expressed by expanding the distribution function of protons to the leading order of $\sigma(x)$,
\begin{equation}
f=f_{0}+\delta f=f_{0}\left( 1-g\sigma /\left( \gamma T\right) \right) ,
\end{equation}%
where $f_{0}$ represents the distribution function in equilibrium, $\delta f$ is the induced fluctuation term,
$\gamma$ is the covariant Lorentz factor, and g is the coupling constant between $\sigma$ field and protons.
From this expansion, the correlators of proton's distribution function are written as
$\left\langle \delta f_{1}...\delta f_{n}\right\rangle _{c} = \left(-\frac{ g }{T}\right)^n \left(\frac{f_{01}...f_{0n}}{\gamma_1...\gamma_n}\right) \left\langle \sigma _{1}...\sigma _{n}\right\rangle _{c}$ $(n=2,3,4,...)$,
where the correlators of $\sigma$ field are obtained from its probability distribution function with cubic
and quartic terms ~\cite{Stephanov:2008qz,Stephanov:2011pb}
\begin{align}
P[\sigma ]=&\exp \left\{ -\Omega \left[ \sigma \right] /T\right\} \notag \\
          =& \exp \left\{-\int d^{3}x\left[ \frac{1}{2}\left( \nabla
\sigma \right) ^{2}+\frac{1}{2}m_{\sigma }^{2}\sigma ^{2}+\frac{\lambda _{3}%
}{3}\sigma ^{3}+\frac{\lambda _{4}}{4}\sigma ^{4}\right]/T\right\}.
\end{align}
The multiplicity fluctuations for protons are obtained by integrating the correlators of proton distribution over the freeze-out surface,
\begin{align}
\left\langle \left( \delta N\right) ^{2}\right\rangle _{c} =& \left( \frac{%
1}{\left( 2\pi \right) ^{3}}\right) ^{2}\prod_{i=1,2}\left( \int \frac{1%
}{E_{i}}d^{3}p_{i}\int_{\Sigma _{i}}p_{i\mu }d\sigma _{i}^{\mu }\right)
\frac{f_{01}f_{02}}{\gamma _{1}\gamma _{2}}\frac{g^{2}}{T^{2}}%
\left\langle \sigma _{1}\sigma _{2}\right\rangle _{c}, \\
\left\langle \left( \delta N\right) ^{3}\right\rangle _{c} =& \left( \frac{%
1}{\left( 2\pi \right) ^{3}}\right) ^{3}\prod\limits_{i=1,2,3}\left(
\int \frac{1}{E_{i}}d^{3}p_{i}\int_{\Sigma _{i}}p_{i\mu }d\sigma _{i}^{\mu
}\right)
\frac{f_{01}f_{02}f_{03}}{\gamma _{1}\gamma _{2}\gamma _{3}}\left(
-1\right) \frac{g^{3}}{T^{3}}\left\langle \sigma _{1}\sigma _{2}\sigma
_{3}\right\rangle _{c}, \\
\left\langle \left( \delta N\right) ^{4}\right\rangle _{c} =& \left( \frac{%
1}{\left( 2\pi \right) ^{3}}\right) ^{4}\prod\limits_{i=1,2,3,4}\left(
\int \frac{1}{E_{i}}d^{3}p_{i}\int_{\Sigma _{i}}p_{i\mu }d\sigma _{i}^{\mu
}\right)
\frac{f_{01}f_{02}f_{03}f_{04}}{\gamma _{1}\gamma _{2}\gamma
_{3}\gamma _{4}}\frac{g^{4}}{T^{4}}\left\langle \sigma _{1}\sigma _{2}\sigma
_{3}\sigma _{4}\right\rangle _{c}.
\end{align}

To calculate the critical fluctuations in equation (3)-(5), we input the information of the freeze-out surface from the hydrodynamic code VISH2+1~\cite{Song:2007fn}. Besides this, several parameters such as $g, \xi, \lambda_3, \lambda_4$ are needed to be input. We set the critical point close to the chemical freeze-out point at 19.6 GeV, and tune these parameters according to the ranges and monotonicity suggested by effective models. For details, one can refer to Ref. \cite{Jiang} for the exact set ups of the parameters.
Note that the critical fluctuations here belong to the static critical fluctuations. In the infinite volume limit of equations (3)-(5), the results given by Stephanov in 2009~\cite{Stephanov:2008qz} can be reproduced.

\section{Results and discussion}

\begin{figure*}[tbp]
\center
\includegraphics[width=2.45 in]{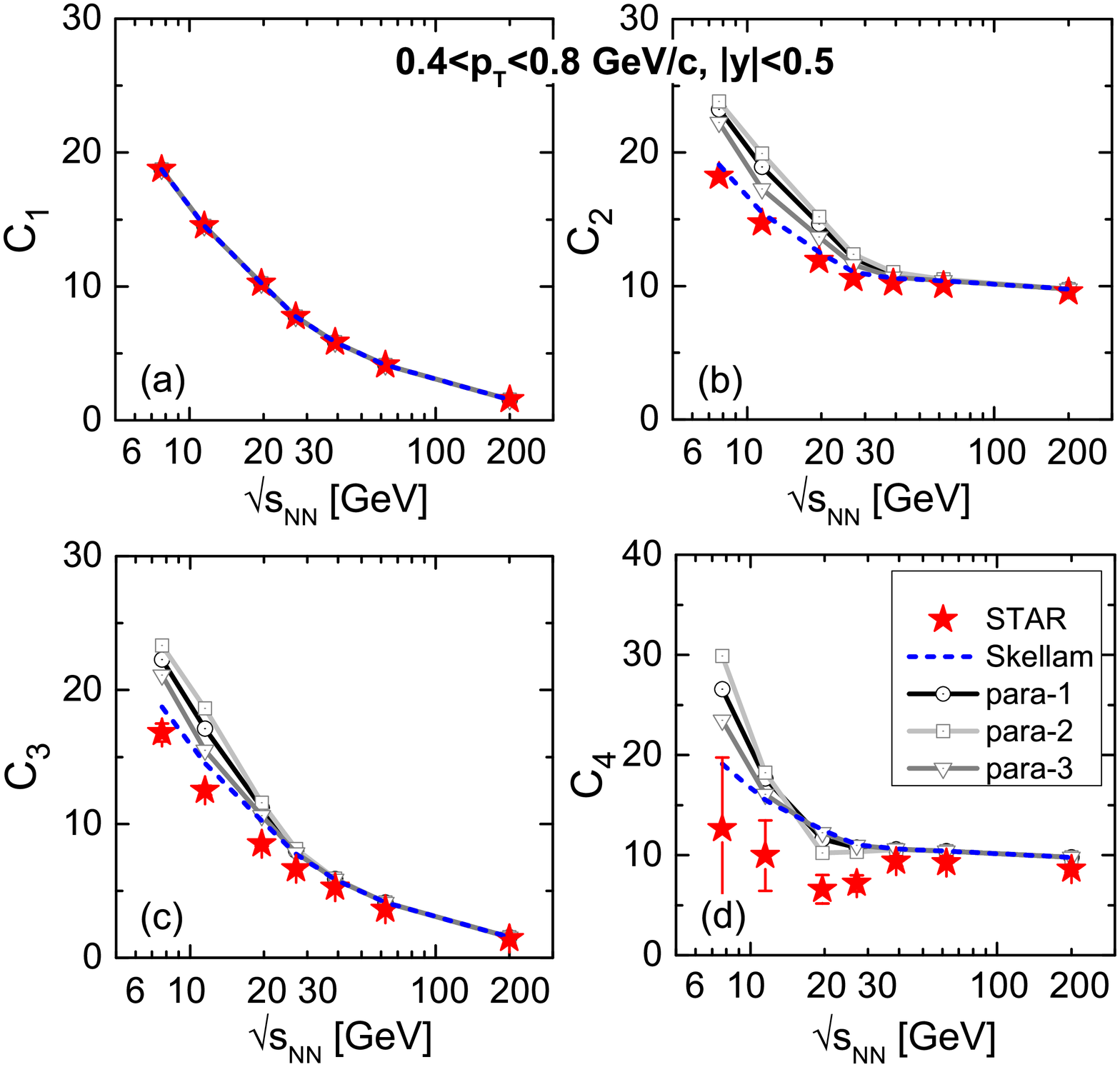} %
\includegraphics[width=2.45 in]{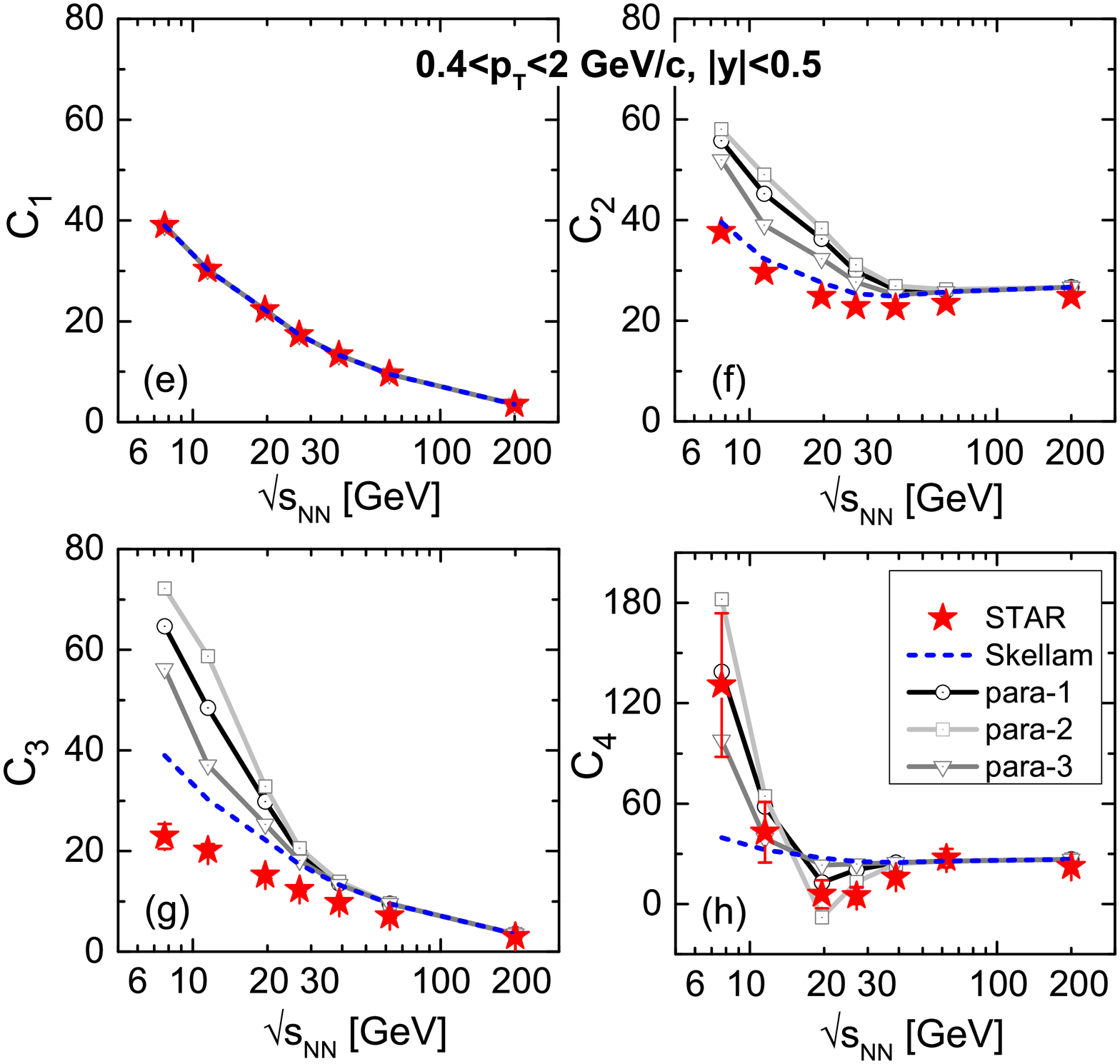}
\caption{Energy dependence of cumulants $C_1-C_4$  for net protons in 0-5\%  Au+Au collisions with Poisson baselines, within $0.4 <p_T< 0.8\ \mathrm{GeV}$ (left panels) and within $0.4 <p_T< 2\ \mathrm{GeV}$ (right panels).  }
\label{cumulant4-ske-0005}

\center
\includegraphics[width=2.7 in]{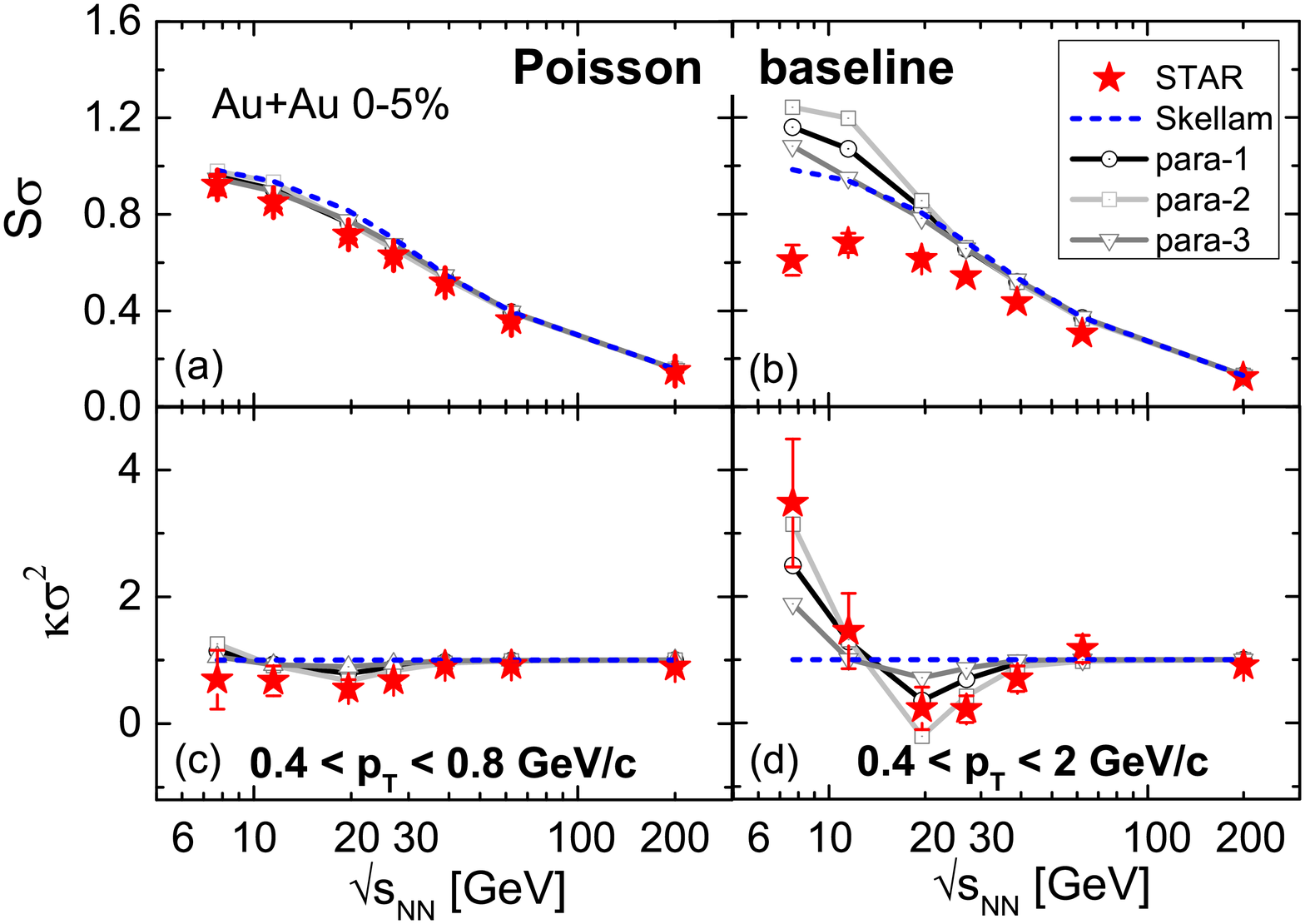}
\caption{Energy dependence of cumulants ratios $S \sigma$ and $\kappa \sigma^2$, for net protons in  0-5\%
Au+Au collisions.}
\label{c-ratio}

\center
\includegraphics[width=4.5 in]{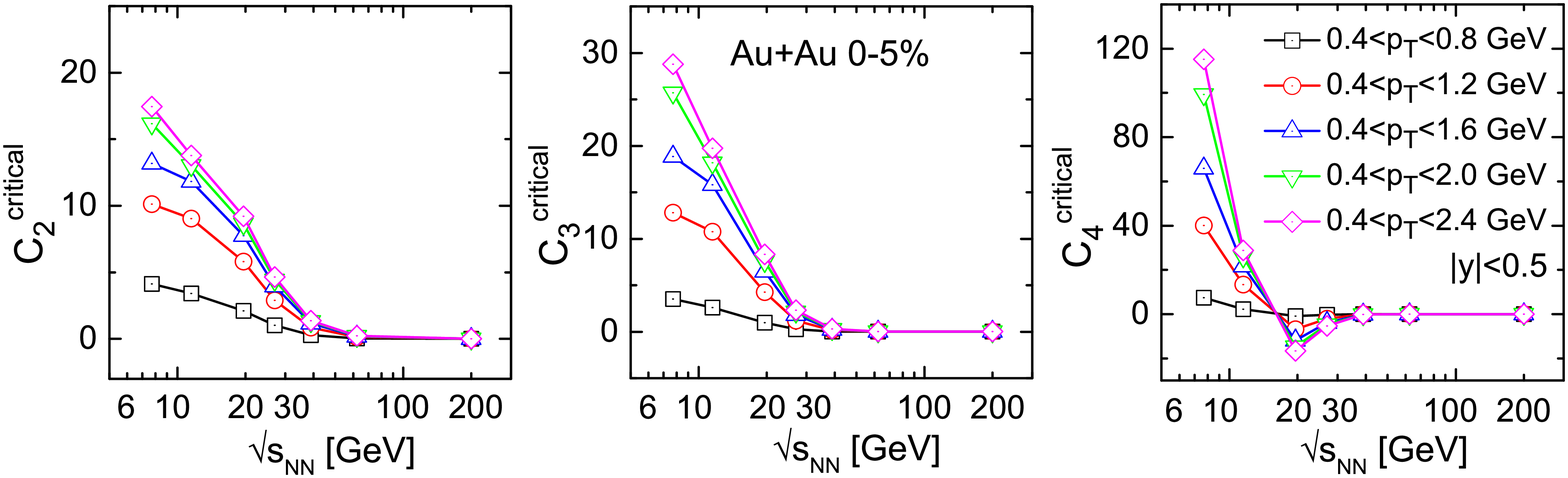}
\caption{$P_T$ acecptance dependence of $C_2^{critical}$, $C_3^{critical}$ and $C_4^{critical}$ for net protons in  0-5\%
Au+Au collisions.}
\label{momentum}
\end{figure*}

Fig.~\ref{cumulant4-ske-0005} presents the results of energy dependent cumulants for net protons at different collision energies and within different $p_T$ ranges, together with Poisson statistical baselines. The theoretical results are the sum of thermal fluctuations and critical fluctuations. Our model calculations present the general trends of the cumulants within different $p_T$ ranges, and can roughly describe the $C_4$ data within error bars. But for $C_2$ and $C_3$, the theoretical calculations deviate from the experimental data which are below the Poisson baselines. This is because the static critical fluctuations provide positive contributions to $C_2$ and $C_3$, a summation with the baselines makes the model calculations deviate further from the experimental data. This problem could not be solved in the framework of static critical fluctuations.

Fig. \ref{c-ratio} presents comparison of model calculations with the experimental data of the cumulants ratios, $S\sigma = C_3/C_2$ and $\kappa \sigma^2 = C_4/C_2$, within different $p_T$ ranges. The experimental data can be roughly described by our model calculation, except $S\sigma$ at lower collision energies within $p_T$ ranges (0.4, 2) GeV, which is caused by the positive contributions of critical fluctuations for $C_2$ and $C_3$.

The experimental cumulants and cumulants ratios also present large enhancements at lower collision energies as the maximum $p_T$ increased from 0.8 to 2 GeV, which can not be described by the statistical baselines. Fig. \ref{momentum} presents our calculation of the $p_T$ acceptance dependence of critical fluctuations of net protons, which shows that the critical fluctuations increase dramatically as the $p_T$ acceptance enlarged. This is because at lower collision energies, the net proton multiplicities increase a lot as the increase of the momentum acceptance, as $\left\langle(\delta N)^n\right\rangle_c \sim N^n$. An increase of the multiplicities leads to the enhancements of cumulants, even though the correlation length is largely reduced here. At higher collision energies, the net proton multiplicities decreases a lot, and the critical fluctuations become negligible.


\section{Summary}
In this proceeding, we summarized the main results in our earlier paper \cite{Jiang}, in which a freeze-out scheme near the QCD critical point for dynamical models is introduced. Based on this freeze-out scheme,
We can describe the energy dependent data of $C_4$ and $\kappa \sigma^2$ within different $p_T$ ranges, and qualitatively describe the momentum acceptance dependent enhancements of cumulants and cumulants ratios at lower collision energies, but always overpredict $C_2$ and $C_3$. A further study of the dynamical critical fluctuations should shed lights on the simultaneous descriptions of different cumulants of net protons in experiments.


\begin{thebibliography}{00}

\bibitem{Stephanov:2008qz}  M.~A.~Stephanov,
Phys.\ Rev.\ Lett.\ \textbf{102}, 032301 (2009).
\bibitem{Stephanov:2011pb}  M.~A.~Stephanov,
Phys.\ Rev.\ Lett.\ \textbf{107}, 052301 (2011).

\bibitem{Aggarwal:2010wy}  M.~M.~Aggarwal \textit{et al.} [STAR
Collaboration],
Phys.\ Rev.\ Lett.\ \textbf{105}, 022302 (2010);
\bibitem{Adamczyk:2013dal}
L.~Adamczyk \textit{et al.} [STAR Collaboration],
Phys.\ Rev.\ Lett.\ \textbf{112}, 032302 (2014).


\bibitem{Luo:2015ewa}  X.~Luo [STAR Collaboration],
PoS CPOD \textbf{2014}, 019 (2014).  


\bibitem{Paech:2003fe}  K.~Paech, H.~Stoecker and A.~Dumitru,
Phys.\ Rev.\ C \textbf{68}, 044907 (2003);  
 M.~Nahrgang, S.~Leupold, C.~Herold and
M.~Bleicher,
Phys.\ Rev.\ C \textbf{84}, 024912 (2011);  
C.~Herold, M.~Nahrgang, I.~Mishustin and
M.~Bleicher,

\bibitem{Jiang}
  L.~Jiang, P.~Li and H.~Song,
  Phys.\ Rev.\ C {\bf 94}, 024918 (2016);
  arXiv:1512.07373 [nucl-th].




\bibitem{Song:2007fn}  H.~Song and U.~W.~Heinz,
Phys.\ Lett.\ B \textbf{658}, 279 (2008);
Phys.\ Rev.\ C \textbf{77}, 064901 (2008). 





\end{thebibliography}
\end{document}